\documentclass[twocolumn,aps,pra,showpacs,superscriptaddress,longbibliography]{revtex4-1}
\usepackage{amsfonts}
\usepackage{amssymb}
\usepackage{mathrsfs}
\usepackage{graphicx}
\usepackage{float}
\usepackage{xcolor}

\begin{document}

\title{Localization and topological transitions in generalized non-Hermitian
SSH models}
\author{X. Q. Sun}
\author{C. S. Liu}
\email{csliu@ysu.edu.cn}
\date{\today }

\affiliation{Hebei Key Laboratory of Microstructural Material Physics,
School of Science, Yanshan University, Qinhuangdao, 066004, China.}

\affiliation{Hebei Key Laboratory of Microstructural Material Physics,
School of Science, Yanshan University, Qinhuangdao, 066004, China.}

\begin{abstract}
We study the localization and topological transitions of the generalized
non-Hermitian SSH models, where the non-Hermiticities are introduced by the
complex quasiperiodic hopping and the nonreciprocal hopping. We elucidate
the universality of the models and how many models can be mapped to them.
Under the open boundary condition, two delocalization transitions are found
due to the competition between the Anderson localization and the boundary
localization from the nontrivial edge states and the non-Hermitian skin
effect. Under the periodic boundary condition, only one delocalization
transition is found due to the disappearance of the non-Hermitian skin
effect. The winding numbers of energy and the Lyapunov exponents in
analytical form are obtained to exactly characterize the two
deloaclizateon transitions. It finds that the delocalization transitions don't
accompany the topological transition. Furthermore, the large on-site
non-Hermiticity and the large nonreciprocal hopping are all detrimental to
the topological transitions. However, the large nonreciprocal hopping
enhances the Anderson localizations. The above analyses are verified by
calculating the energy gap and the inverse of the participation ratio
numerically.
\end{abstract}

\maketitle

\section{Introduction}

As the topological edge states are protected by the symmetry, the
topological phases are expected to be immune to the perturbations of
disorders \cite{RevModPhys.82.1959, RevModPhys.82.3045, RevModPhys.83.1057}.
Nevertheless, the strong disorders induce the destructive interference of
scattered waves and lead to the localization of all the states \cite%
{PhysRev.109.1492, Abrahams}. The presence of the Anderson localized phase
accompanies the disappearance of the topological nontrivial phase. The
disorders connect the two unrelated quantum states which give rise to the
topological Anderson insulators \cite{PhysRevLett.102.136806,
PhysRevLett.103.196805, PhysRevB.80.165316}. Recently, the ability to
engineer non-Hermitian Hamiltonians and the related observations of
unconventional topological edge modes attract a great interest to extend the
topological band theory to non-Hermitian systems \cite{RevModPhys.93.015005}%
. The interplays between the topology and non-Hermiticity result in a
plethora of exotic phenomena that have no Hermitian counterparts, e.g., the
Weyl exceptional ring \cite{PhysRevLett.118.045701}, the point gap \cite%
{PhysRevX.8.031079}. In particular, the realization of the nonreciprocal
hopping induces the non-Hermitian skin effect. Comparing to the Hermitian
case where the zero-energy states are localized at the boundary only, the
non-Hermitian skin effect results in the asymmetrical localization where not
only the nontrivial edge states, but also all the trivial states are
localized at the boundary. The competition of the boundary localization and
bulk localization must lead to the delocalization transitions. In the
presence of the non-Hermitian skin effect, a characteristic signature is the
breakdown of the conventional bulk-boundary correspondence \cite%
{RevModPhys.93.015005, PhysRevB.97.121401, PhysRevLett.121.026808,
PhysRevLett.121.086803, PhysRevLett.121.136802, PhysRevLett.123.066404,
PhysRevLett.123.170401, PhysRevB.99.201103, PhysRevLett.123.246801,
PhysRevLett.124.056802, PhysRevLett.125.126402, PhysRevB.99.081103,
PhysRevB.101.020201, PhysRevB.102.085151, PhysRevLett.125.126402,
PhysRevLett.125.186802, PhysRevLett.125.226402, PhysRevB.99.081103, Han_2021}%
. The question arises as to whether or not the delocalization transitions
accompany the topological transitions. Where are the phase boundaries and
how the non-Hermitian skin effect modifies the phase transitions?

The Anderson localizations are generally studied with the Aubry-Andr\'{e}%
-Harper (AAH) model which is the periodic lattice modulated by
quasiperiodic potential. Quasicrystals constitute an intermediate phase
between fully periodic lattices and fully disordered media, showing a
long-range order without periodicity \cite{AIPS.3.133, Harper_1955}. This
solves the problems of reliably controlling disorders in solid-state systems
where the Anderson localization could only be studied indirectly or through
numerical simulations. As the AAH model can be mapped to the lattice version
of the two-dimensional integer Hall effect problem, it gives the opportunity
to study the Anderson localization and topological properties simultaneously
\cite{PhysRevLett.108.220401, PhysRevLett.109.106402}. The exact
localization-delocalization transition can be obtained by the duality
transformation between real and momentum spaces \cite%
{PhysRevLett.104.070601, PhysRevB.83.075105, PhysRevLett.114.146601,
PhysRevB.93.184204, PhysRevB.96.085119, PhysRevB.101.064203,
PhysRevB.103.014203, PhysRevLett.123.025301, PhysRevLett.123.070405,
PhysRevLett.125.060401, 10.21468/SciPostPhys.12.1.027}.

When the non-Hermiticities are further introduced to the AAH model by the
complex on-site quasiperiodic potentials, it was found that the self-dual
symmetry still determines the transitions from topological nontrivial phase
to localized phase \cite{PhysRevLett.122.237601, PhysRevResearch.2.033052,
PhysRevB.100.054301, PhysRevB.102.024205, PhysRevB.103.214202,
PhysRevB.104.024201}. The quantum phase transitions still have the
topological nature characterized by the winding numbers. However, when the
non-Hermiticity is from nonreciprocal hopping, the induced non-Hermitian
skin effect leads to the asymmetric localized states \cite%
{PhysRevB.100.054301, PhysRevB.103.214202, PhysRevB.104.024201}. It find
that the Anderson localization is not necessarily in accordance with the
topological phase transitions. A dip is found in the quasiperiodic potential
dependence of the averaged inverse participation ratios which indicates the
delocalization transition \cite{PhysRevB.100.054301}. The transition point
can be exactly proved by the skin effect rescaling.

Up to now, the studies of the non-Hermitian topological Anderson transitions
mainly focus on the AAH models \cite{PhysRevB.100.054301,
PhysRevResearch.2.033052, PhysRevB.103.014203, PhysRevB.104.024201,
PhysRevA.103.033325}. The Su-Schrieffer-Heeger (SSH) model for polyacetylene
is the simplest one-dimensional topological insulator with the chirality
symmetry \cite{PhysRevLett.42.1698}. As many non-Hermitian models can be
mapped to various non-Hermitian versions of the extended SSH models \cite%
{Han_2021}, they have become the typical models to study the non-Hermitian
effects. In particular, in the presence of the non-Hermitian skin effect due
to the nonreciprocal hopping, the generalized Brillouin zone of the
non-Hermitian SSH model is proposed to recover the bulk-boundary
correspondence \cite{ PhysRevB.97.121401, PhysRevLett.121.026808,
PhysRevLett.121.086803, PhysRevLett.121.136802, PhysRevLett.123.066404,
PhysRevLett.123.170401, PhysRevB.99.201103, PhysRevLett.123.246801,
PhysRevLett.124.056802, PhysRevLett.125.126402, PhysRevB.99.081103,
PhysRevB.101.020201, PhysRevB.102.085151, PhysRevLett.125.126402,
PhysRevLett.125.186802, PhysRevLett.125.226402, PhysRevB.99.081103, Han_2021}%
. %The
%generalized Brillouin zone (GBZ) is proposed to recover the bulk-boundary
%\cite{ PhysRevB.97.121401, PhysRevLett.121.026808,
%PhysRevLett.121.086803, PhysRevLett.121.136802, PhysRevLett.123.066404,
%PhysRevLett.123.170401, PhysRevB.99.201103, PhysRevLett.123.246801,
%PhysRevLett.124.056802, PhysRevLett.125.126402, PhysRevB.99.081103,
%PhysRevB.101.020201, PhysRevB.102.085151, PhysRevLett.125.126402,
%PhysRevLett.125.186802, PhysRevLett.125.226402}.
%Alternatively, inspired by
%the asymmetric coupling equivalent to the imaginary gauge field appling to
%the lattice, the counterpart of the non-Hermitian SSH model is
%constructed by eliminating the effective imaginary gauge field and the
%induced non-Hermitian skin effect \cite{PhysRevB.99.081103, Han_2021}. As
%the results, the topological invariants of the non-Hermitian model can be
%defined in conventional Brillouin zone (BZ).
%The non-Hermitian skin effect localizes all states at boundary whether zero-energy edge states or bulk states. However, the applied disorders on the model result in the Anderson localization on the bulk.
%The localizations of the model are expected to be the competition between
%the boundary localization and the bulk localization.
The influence of the non-Hermitian skin effect on disordered SSH model is
explored and the result shows that the non-Hermiticity can enhance the
topological phase against disorders by increasing bulk gaps \cite%
{CPMA.63.267062}. The onset of a mobility edge and the topological phase
transition in the disordered SSH chain connected to two external baths was
investigated the Lindblad equation method \cite{2022arXiv221010856}. The
localization/delocalization of the disordered chain can be recovered by the
scaling properties of the nonequilibrium stationary current. However, the
winding numbers and Lyapunov exponents in analytical forms are still lacking
to characterize the relationship between the topological transitions and
Anderson transitions.

In this paper, we study the non-Hermitian generalized SSH model, where the
non-Hermiticities are introduced by both the complex quasiperiodic
modulation hopping and the nonreciprocal hopping. The main results of this
article include the following: (i) We find the equivalence and the
universality of the models connected by the similarity transformation.
%(ii)
%We find the effective imaginary gauge field which causes the bulk states and
%edge states exponential localized at the boundaries.
(ii) The winding numbers and the Lyapunov exponents in analytic form are
obtained to characterize the topological transition and localization
transition. (iii) We clarify the origin of the two delocalization
transitions. (iv) The influences of non-Hermiticities on the localizations
and topological phase transitions are given.

The rest of paper is organized as follows: In Sec. \ref{model} we present
the Hamiltonian of the generalized non-Hermitian SSH model and its various
equivalent models. We explain the equivalence of the hopping terms and
elucidate why the complex quasiperiodic modulation hopping can be applied in
any of the terms. We also clarify how the nonreciprocal hopping can be
transformed from one form to another by the similarity transformations. %In
%Sec. \ref{Non-Hermitian skin effect}, we analysis how the effective
%imaginary gauge field induce non-Hermitian skin effect and how to eliminate
%it to obtain its topological equivalent model without non-Hermitian skin
%effect.
The winding numbers and the Lyapunov exponents are derived in Sec. \ref{The
topological transition and the Anderson transition} to characterize the
topological transition and the Anderson transition. In Sec. \ref{Phase
diagram}, the phase diagram of the model is obtained by calculating the
inverse of participation ratio. A brief summary is presented in Sec. \ref%
{summary}.

\section{Model}

\label{model}

We consider the non-Hermitian SSH model [pictorially shown in Fig. \ref{Non
Hermitian SSH model} (a)] \cite{PhysRevLett.42.1698, PhysRevB.102.161101,
Jangjan:2020vu}. The Hamiltonian is described by
\begin{eqnarray}
H\! &=&\Psi ^{\dag }h\!\Psi =\sum\limits_{n}\left[ t_{1,n}a_{n}^{\dagger
}b_{n}+t_{1,n}b_{n}^{\dagger }a_{n}\right.
\label{the non-Hermitian SSH model_real} \\
&&+\left( t_{2}-\gamma _{2}\right) a_{n+1}^{\dagger }b_{n}+\left(
t_{2}+\gamma _{2}\right) b_{n}^{\dagger }a_{n+1}  \nonumber \\
&&\left. +\left( t_{3}-\gamma _{3}\right) a_{n}^{\dagger }b_{n+1}+\left(
t_{3}+\gamma _{3}\right) b_{n+1}^{\dagger }a_{n}\right] ,  \nonumber
\end{eqnarray}%
where Nambu operator $\Psi ^{\dag }=\left( a_{1}^{\dag },b_{1}^{\dag },\cdot
\cdot \cdot ,a_{L}^{\dag },b_{L}^{\dag }\right) $. $a_{n}^{\dagger }$ and $%
b_{n}^{\dagger }$\ ($a_{n}$ and $b_{n}$) are the creation (annihilation)
operators of a spinless fermion at lattice site $n$. $t_{1,n}$, $t_{2}$ and $%
t_{3}$ are the hopping amplitudes. The parameters $\gamma _{2}$ and $\gamma
_{3}$ induce the asymmetric coupling. The model in Eq. (\ref{the
non-Hermitian SSH model_real}) has chiral symmetry $\mathcal{S}h\mathcal{S}%
^{-1}=-h$, with operator $\mathcal{S}=\text{diag}\left( 1,-1,\cdot \cdot
\cdot ,1,-1\cdot \cdot \cdot \right) $. The chiral symmetry ensures the
eigenvalues ($E,-E$) pairs.

\begin{figure}[tbp]
\begin{center}
\hspace*{-0.1cm} \includegraphics[scale=0.26]{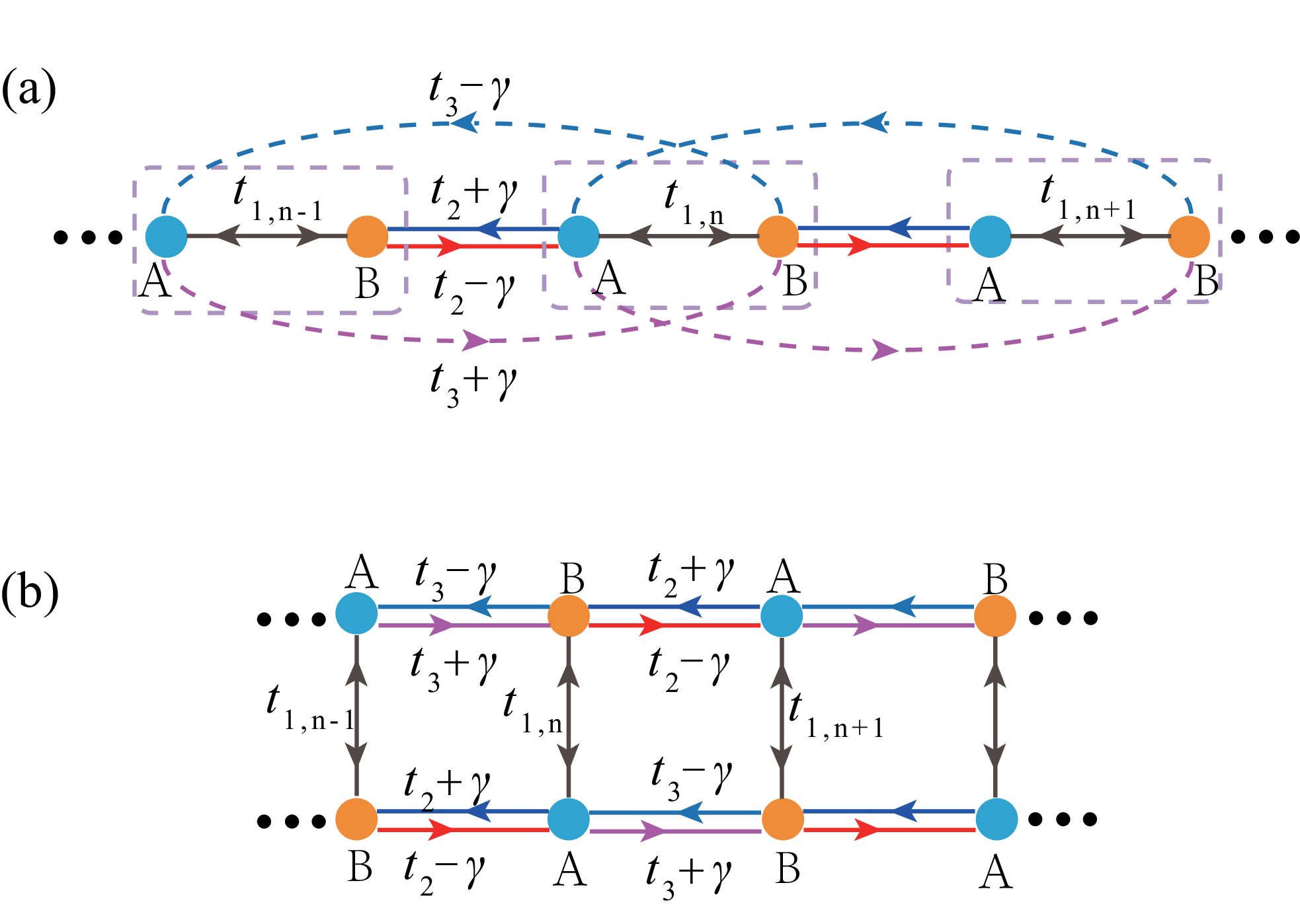}
\end{center}
\caption{(a) Non-Hermitian SSH model and (b) its equivalent two-leg ladder
model. }
\label{Non Hermitian SSH model}
\end{figure}

The hopping amplitude $t_{1,n}$ is set to be site dependent, i.e.
\begin{equation}
t_{1,n}=2V\mathrm{cos}(2\pi \beta n+i\alpha ),
\label{quasiperdoic potential}
\end{equation}%
with the strength $V$. $\alpha $ depicts the
non-Hermiticity of the quasiperiodic hopping. $\beta $ is an irrational number which is used to
characterize the quasiperiodicity. The irrational number is usually taken by the value of the inverse of golden
ratio [$\beta =(\sqrt{5}-1)/2$] which in practice approximated by rational
numbers $\beta =F_{n}/F_{n+1}$ with the $n$th Fibonacci number $F_{n}$. The $%
n$th Fibonacci number $F_{n}=F_{n-1}+F_{n-2}$ and the $1$th and $2$th
Fibonacci numbers are assumed to be $F_{1}=1$ and $F_{2}=1$ respectively.
In numerical simulations, a finite (yet arbitrarily large)
number of lattice sites $L=F_{n+1}$ is assumed usually to
eliminate the domain wall of the PBC and to reduce the finite-size effects.
The quasiperiodic term $t_{1,n}
$ acts as the disorders and induces localizations of the states.

It should be emphasized that $t_{1,n}$, $t_2$ and $t_3$ are equivalent and the quasiperiodicity can
be set to any of the terms. The reason is the model in Fig. \ref{Non
Hermitian SSH model} (a) is equivalent to the two-leg ladder model
[pictorially shown in Fig. \ref{Non Hermitian SSH model} (b)]. Exchanging $%
t_{1,n}$ and $t_{2}$ terms, the AB lattice of the model becomes merely the
BA lattice which corresponds to relabeling $b_{n}\rightarrow b_{n+1}$. Similarly, the model
remains unchanged when exchanging $t_{2}$ and $t_{3}$ terms.

The introduction of parameters $\gamma _{2}$ and $\gamma _{3}$ do not lose
the universality since the asymmetry of hopping $t_{2}\pm \gamma _{2} = \tau_{\pm} = \bar{t}%
_{2}e^{\pm \phi _{2}}$ and $t_{3}\pm \gamma _{3}=\bar{t}_{3}e^{\pm \phi
_{3}} $ can be shifted from one term to another with the similarity
transformations under the open boundary condition (OBC) \cite{Han_2021}.
After the similarity transformations, we obtain the equivalent models.
%Although the eigenvalues
%of original models are different from that of its equivalent models, they
%have the same zero energy points in the energy spectra. So the original
%model and its equivalent models are topological equivalent which can be
%found in the Appendix.
For example, under the OBC, the Hamiltonian of this model in Eq. (\ref{the
non-Hermitian SSH model_real}) can be transformed into%
\begin{eqnarray}
\hat{H} &=&\hat{\Psi}^{\dag }\hat{h}\!\hat{\Psi}=\sum\limits_{n}\left[
t_{1,n}\hat{a}_{n}^{\dagger }\hat{b}_{n}+t_{1,n}\hat{b}_{n}^{\dagger }\hat{a}%
_{n}\right.  \label{non-Hermiticity transformation t2 and t3} \\
&&+\left( t_{2}-\gamma _{2}\right) r_{1}^{-1}\hat{a}_{n+1}^{\dagger }\hat{b}%
_{n}+\left( t_{2}+\gamma _{2}\right) r_{1}\hat{b}_{n}^{\dagger }\hat{a}_{n+1}
\nonumber \\
&&\left. +\left( t_{3}-\gamma _{3}\right) r_{1}\hat{a}_{n}^{\dagger }\hat{b}%
_{n+1}+\left( t_{3}+\gamma _{3}\right) r_{1}^{-1}\hat{b}_{n+1}^{\dagger }%
\hat{a}_{n}\right] ,  \nonumber
\end{eqnarray}%
by the similarity transformation $\hat{h}=S_{1}^{-1}hS_{1}$ and $\hat{\Psi}%
=S_{1}^{-1}\Psi $ with
\begin{equation}
S_{1}=\text{diag}\{r_{1},r_{1},r_{1}^{2},r_{1}^{2},\cdots
,r_{1}^{L/2},r_{1}^{L/2}\}.  \label{the transformation matrix S1}
\end{equation}%
When $r_{1}=e^{\phi _{3}}$, the non-Hermiticity in $t_{3}$ term is killed
and the non-Hermiticity is transformed into $t_{2}$ term completely.
However, when $r_{1}=e^{-\phi _{2}}$, the non-Hermiticity in $t_{2}$ term is
eliminated and the non-Hermiticity remains in $t_{3}$ term. In such case $%
r_{1}=e^{-\phi _{2}}$, further doing the transformation $\check{h}%
=S_{2}^{-1}\hat{h}S_{2}$ and $\check{\Psi}=S_{2}^{-1}\hat{\Psi}$ with
\begin{equation}
S_{2}=\text{diag}\{1,r_{2},r_{2},r_{2}^{2},r_{2}^{2},\cdots
,r_{2}^{L/2-1},r_{2}^{L/2-1},r_{2}^{L/2}\},
\label{the transformation matrix S2}
\end{equation}%
and $r_{2}=e^{-\left( \phi _{2}+\phi _{3}\right) /2}$, the model in Eq. (\ref%
{non-Hermiticity transformation t2 and t3}) becomes%
\begin{eqnarray}
\check{H} &=&\check{\Psi}^{\dag }\check{h}\!\check{\Psi}=\sum\limits_{n}%
\left[ {t}_{1,n}r_{2}^{-1}\check{a}_{n}^{\dagger }\check{b}_{n}+{t}%
_{1,n}r_{2}\check{b}_{n}^{\dagger }\check{a}_{n}\right.
\label{non-Hermiticity transformation t3 to t1} \\
&&\left. +\bar{t}_{2}\check{a}_{n+1}^{\dagger }\check{b}_{n}+\bar{t}_{2}%
\check{b}_{n}^{\dagger }\check{a}_{n+1}+\bar{t}_{3}\check{a}_{n}^{\dagger }%
\check{b}_{n+1}+\bar{t}_{3}\check{b}_{n+1}^{\dagger }\check{a}_{n}\right] .
\nonumber
\end{eqnarray}%
The non-Hermiticity in $t_{3}$ term is transformed into $t_{1}$ term.

It is easy to define the unity operator $U$ which transforms the Numb
creation operator $\Psi ^{\dag }$ in Eq. (\ref{the non-Hermitian SSH
model_real}) to
\begin{eqnarray*}
\Phi ^{\dag } &=&\Psi ^{\dag }U=\left( a_{1}^{\dag },b_{1}^{\dag },\cdot
\cdot \cdot ,a_{L}^{\dag },b_{L}^{\dag }\right) U \\
&=&\left( a_{1}^{\dag },a_{2}^{\dag },\cdot \cdot \cdot ,a_{L}^{\dag
},-ib_{1}^{\dag },-ib_{2}^{\dag },\cdot \cdot \cdot ,-ib_{L}^{\dag }\right)
\\
&=&\left( u^{\dag },v^{\dag }\right) .
\end{eqnarray*}%
With the unity operator $U$, the Hamiltonian $h$ in Eq. (\ref{the
non-Hermitian SSH model_real}) under the PBC\ can be transformed into two
block off-diagonal form%
\begin{equation}
\tilde{h}=U^{\dag }hU=-i\tilde{h}_{+}\sigma _{+}+i\tilde{h}_{-}\sigma _{-}
\label{the block off-diagonal Hamiltonian}
\end{equation}%
here
\begin{eqnarray*}
\tilde{h}_{+} &=&\left(
\begin{array}{cccc}
t_{1,n} & t_{3}-\gamma _{3} &  & t_{2}-\gamma _{2} \\
t_{2}-\gamma _{2} & t_{1,n} & \ddots &  \\
& \ddots & t_{1,n} & t_{3}-\gamma _{3} \\
t_{3}-\gamma _{3} &  & t_{2}-\gamma _{2} & t_{1,n}%
\end{array}%
\right) , \\
\tilde{h}_{-} &=&\left(
\begin{array}{cccc}
t_{1,n} & t_{2}+\gamma _{2} &  & t_{3}+\gamma _{3} \\
t_{3}+\gamma _{3} & t_{1,n} & \ddots &  \\
& \ddots & t_{1,n} & t_{2}+\gamma _{2} \\
t_{2}+\gamma _{2} &  & t_{3}+\gamma _{3} & t_{1,n}%
\end{array}%
\right)
\end{eqnarray*}%
and $\sigma _{\pm }=\left( \sigma _{x}\pm i\sigma _{y}\right) /2$. Under the
OBC, there aren't non-diagonal elements in the matrix $\tilde{h}_{+}$ and $%
\tilde{h}_{-}$. In the new Numb operator $\Phi$, the Hamiltonian in Eq. (\ref{the
non-Hermitian SSH model_real}) becomes $H\!=\Phi ^{\dag }\tilde{h}\!\Phi $
and the eigenfunction is given by $\tilde{h}\!\Phi =E_{n}\Phi $.

With the block off-diagonal form, the eigenvalue problem turns into%
\begin{eqnarray*}
-i\tilde{h}_{+}v_{n} &=&E_{n}u_{n}\text{, } \\
i\tilde{h}_{-}u_{n} &=&E_{n}v_{n}.
\end{eqnarray*}%
We then obtain%
\begin{eqnarray}
\tilde{h}_{+}\tilde{h}_{-}u_{n} &=&E_{n}^{2}u_{n},  \nonumber \\
\tilde{h}_{-}\tilde{h}_{+}v_{n} &=&E_{n}^{2}v_{n}.
\label{the eigenvalue problem}
\end{eqnarray}%
$u_{n}$ and $v_{n}$ can be treated as the single-particle states, and have
the same properties. We find the zero-energy states by sovling the Eq. (\ref%
{the eigenvalue problem}). The merit is the dimmension of $\tilde{h}_{+}%
\tilde{h}_{-}$ or $\tilde{h}_{-}\tilde{h}_{+}$ is equal to half of that of $%
h $ in Eq. (\ref{the non-Hermitian SSH model_real}).

Replacing $t_{2}\rightarrow -t-\Delta $, $t_{3}\rightarrow -t+\Delta $, $%
\gamma _{2}=\gamma _{3}=\gamma $ and $t_{1,n}\rightarrow V_{j}$, the
non-Hermitian SSH Hamiltonian in Eq. (\ref{the block off-diagonal
Hamiltonian}) is mapped to non-Hermitian AAH models with p-wave pairing \cite%
{PhysRevB.103.214202}%
\begin{eqnarray}
H &=&\sum_{j}\left[ -\left( t+\gamma \right) c_{j}^{\dagger }c_{j+1}-\left(
t-\gamma \right) c_{j+1}^{\dagger }c_{j}\right.  \nonumber \\
&&\left. +\Delta c_{j}c_{j+1}+\Delta c_{j+1}^{\dag }c_{j}^{\dag }\right]
+\sum_{j}V_{j}c_{j}^{\dagger }c_{j}  \nonumber \\
&=&\tilde{\Psi}_{L}^{\dag }\tilde{h}\!\tilde{\Psi}_{R},
\label{non-Hermitian AAH models with p-wave pairing}
\end{eqnarray}%
in the Majorana fermion representation. $c_{j}^{\dagger }$ is the creation
operator of a spinless fermion at lattice site $j$ and $t$ is the hopping
amplitude and set as the unit energy ($t=1$). $\Delta $ is the p-wave
pairing amplitude. The Majorana fermion operator%
\[
\tilde{\Psi}_{L}^{\dag }=\left( \nu _{1}^{A},\nu _{2}^{A}\cdot \cdot \cdot
\nu _{L}^{A},\nu _{1}^{B},\nu _{2}^{B}\cdot \cdot \cdot \nu _{L}^{B}\right)
,
\]%
where $\nu _{j}^{A}\equiv c_{j}+c_{j}^{\dagger }$ and $\nu _{j}^{B}\equiv
i(c_{j}-c_{j}^{\dagger })$ are operators of two Majorana fermions belonging
to one physical site. They satisfy relations $(\nu _{j}^{\kappa })^{\dagger
}=\nu _{j}^{\kappa }$ and $\{\nu _{j}^{\kappa },\nu _{k}^{\lambda
}\}=2\delta _{jk}\delta _{\kappa \lambda }$ ($\kappa ,\lambda =A,B$) \cite%
{Kitaev_2001, PhysRevLett.110.176403}. In the case of $\gamma =0$,\
localization and topological phase transitions of model (\ref{non-Hermitian
AAH models with p-wave pairing}) have been studied in Ref. \cite%
{PhysRevB.103.214202}.% by analyzing the $\mathcal{PT}$ symmetry
%breaking, winding numbers of energy spectra, localization and fractal
%dimensions of states, and fate of Majorana fermions.

\section{ The topological transition and the Anderson transition}

\label{The topological transition and the Anderson transition}

%Topological materials are characterized by robust boundary states dictated by the bulk topological invariants.
%The topological phase is
%expected to be against to perturbations of disorder due to the protection of
%the symmetry. However, the arbitrarily small disorders could induce the destructive
%interference which induces the Anderson localization.

\subsection{Winding number}

Considering the periodicity of AB lattice and the modulated quasiperiodic
potential, the topological transitions can in principle be discussed in GBZ
with the technique proposed in Ref. \cite{PhysRevB.97.121401,
PhysRevLett.121.026808, PhysRevLett.121.086803, PhysRevLett.121.136802,
PhysRevLett.123.066404, PhysRevLett.123.170401, PhysRevB.99.201103,
PhysRevLett.123.246801, PhysRevLett.124.056802, PhysRevLett.125.126402,
PhysRevB.99.081103, PhysRevB.101.020201, PhysRevB.102.085151,
PhysRevLett.125.126402, PhysRevLett.125.186802, PhysRevLett.125.226402}. The
topological transition can also be studied with its topological equivalent
model in conventional BZ introduced in Ref. \cite{Han_2021}. However, when
the quasiperiodic potential in Eq. (\ref{quasiperdoic potential}) is used to
depict the disorder, a large order of Fibonacci number must be used to
ensure the long term disorder which results in the complexity of the
discussion. Alternatively, a phase $\delta $ is usually added in the complex
quasiperiodic potential
\[
V_{j}=2V\mathrm{cos}(2\pi \beta j+i\alpha +\delta /L),
\]%
to introduce an additional dimension. The winding number of energy is often
used to characterize the topological nontrivial phase and defined as
\[
\nu =\lim\limits_{L\rightarrow \infty }\oint \frac{d\delta }{2\pi i}\partial
_{\delta }\left[ \ln \det \left( H\right) \right] ,
\]%
here the Hamiltonian $H$ is defined under the PBC \cite{PhysRevX.8.031079}. $%
\nu $ can be used to depict how the complex spectral trajectory $E$
encircles a base energy $0$ in the complex energy plane, when $\delta $
changes from 0 to $2\pi $.

With the block off-diagonal Hamiltonian in Eq. (\ref{the block off-diagonal
Hamiltonian}), the winding number is given by \cite{PhysRevB.83.085426,
PhysRevX.4.011006}
\[
\nu =\lim\limits_{L\rightarrow \infty }\oint \frac{d\delta }{2\pi i}\partial
_{\delta }\left[ \ln \det \left( \tilde{h}_{+}\tilde{h}_{-}\right) \right] ,
\]%
which is the summation of winding numbers of winding vectors $\det \left(
\tilde{h}_{+}\right) $ and $\det \left( \tilde{h}_{-}\right) $ in the
complex energy plane. In the large limit $L\rightarrow \infty $, $\det
\left( \tilde{h}_{+}\right) $ and $\det \left( \tilde{h}_{-}\right) $ have
the analytical forms which are given by
\begin{equation}
\mathrm{\det }(\tilde{h}_{\pm })=-\tau _{\pm }^{L}+(-1)^{L+1}V^{L}e^{L\alpha
-i\delta }+P_{\pm },  \label{the PBC determint}
\end{equation}%
where
\begin{equation}
\tau _{\pm }=\max \left( \bar{t}_{2}e^{\pm \phi _{2}},\bar{t}_{3}e^{\pm \phi
_{3}}\right) \text{ },  \label{The parameter tau_pm}
\end{equation}%
and $P_{\pm }=[\mathrm{max}(\bar{t}_{2}\bar{t}_{3}e^{\pm \left( \phi
_{2}+\phi _{3}\right) },V)]^{L}.$ The detailed calculations can consult the
Refs \cite{PhysRevLett.110.176403, PhysRevB.100.054301, PhysRevB.103.014201,
PhysRevB.103.214202}. $P_{\pm }$ can be neglected under the PBC.
%Under the OBC however, $P_{\pm }$ is
%important and can not be neglected. As shown below, $P_{\pm }$ is important
%to understand the Anderson localization transition under the OBC.
The winding number is given by $\nu =\nu _{+}+\nu _{-}$,
\begin{equation}
\nu _{\pm }=\Theta (Ve^{\alpha }-\tau _{\pm })\text{ },
\label{winding number}
\end{equation}%
here $\Theta (x)$ is the step function. The topological transition points
are given by $V_{\text{T},\pm }=\tau _{\pm }e^{-\alpha }$ which indicates
that the onsite non-Hermiticity $\alpha $ is always negative correlation
with topological transition. Without losing the generality, it assumes that $%
\bar{t}_{2}e^{\phi _{2}}>\bar{t}_{3}e^{\phi _{3}}$ and $\bar{t}_{2}e^{-\phi
_{2}}>\bar{t}_{3}e^{-\phi _{3}}$. We get the parameter $\tau _{\pm }=\bar{t}%
_{2}e^{\pm \phi _{2}}=t_{2}\pm \gamma _{2}$. With increasing the disorder
intensity $V$, the nontrivial edge state is destoryed when $V=\tau
_{-}e^{-\alpha }$. Further increasing the disorder intensity $V$, the system
is still in the trivial topological phase. So the topological transition
point is $V_{\text{T}}=\tau _{-}e^{-\alpha }$ which indicates that the
nonreciprocal parameter $\gamma _{2}$ is negative correlation with
topological transition.

\subsection{Lyapunov exponent}

Under the weak disorder perturbations, the energy-band structure remains
unchanged. For the topological nontrivial Hermitian system, %two different
%mechanisms govern the localization.
the nontrivial edge states are almost localized at several sites while all
the trivial states are Bloch waves and extend in the bulk. In such case, the
localization is boundary localization which is from the nontrivial edge
states. In the strong disorder case however, the energy-band structure is
broken. The localization is determined by Anderson localization where all
the states localize in the bulk. The nonreciprocal hopping induces the
non-Hermitian skin effect and the asymmetrical (boundary) localization. The
strong disorders also break the asymmetrical localization and lead to the
Anderson localization. Accompanying the transition from boundary to Anderson
localization, a delocalization must occur.

The transition from extended to localized phases can be clarified by the
Lyapunov exponent $\eta $ which is the inverse of localization length. In
the localized phase, the wave function is exponentially decayed in the
localization center $n_{0}$, i.e.$\ \mathrm{exp}\left( -\eta \left\vert
n-n_{0}\right\vert \right) $. Concerning various localizations, four kinds
of the Lyapunov exponents are needed to depict the localizations of the
model.

%For the non-Hermitian system, the non-Hermitian skin effect drives all the states to the boundary. However, the strong disorders result in Anderson localization.

%The two kinds of competition lead to the delocalization phase.
%The non-Hermiticities modify not only the boundary localization but also bulk localization.
%The localization is firstly
%determined by \cite{PhysRevB.88.165111}.%also induces boundary localization, the
%localizations of the topological materials are expected to enhance the boundary localization.
%One is
%from the localization of the nontrivial edge state of the model
%discussed in Ref. which causes the particles localized near the boundary. The other is the Thouless
%mechanism which is the destructive interference of disorders in the whole
%material given in Ref. \cite{Thouless_1972}. This effect
%leads to the particles localization in the bulk which causes the edge states
%delocalization. The localization of the system is competition of
%the two mechanisms.

The first Lyapunov exponent is related to the nontrivial topological state
whose definition is based on the transfer matrix approach \cite%
{PhysRevB.88.165111}. If the system under OBC hosts a topological nontrivial
state, the block off-diagonal Hamiltonian in Eq. (\ref{the block
off-diagonal Hamiltonian}) has the zero-energy mode which should satisfy $%
\tilde{h}_{+}v_{n}=0$. In transfer matrix form, the equation can be
rewritten as%
\[
\left(
\begin{array}{c}
v_{j+1} \\
v_{j}%
\end{array}%
\right) =T_{j}\left(
\begin{array}{c}
v_{j} \\
v_{j-1}%
\end{array}%
\right) ,
\]%
with
\[
T_{j}=\left(
\begin{array}{cc}
-\frac{V_{j}}{t_{3}-\gamma _{3}} & -\frac{t_{2}-\gamma _{2}}{t_{3}-\gamma
_{3}} \\
1 & 0%
\end{array}%
\right) .
\]%
The transfer matrix of the whole system is given by $T=\Pi _{j=1}^{L}T_{j}$
with the two eigenvalues $\lambda _{1}$ and $\lambda _{2}$. Supposing $%
|\lambda _{1}|<|\lambda _{2}|$, the Lyapunov exponent of the edge states is
defined as
\begin{equation}
\eta _{1}=-\lim_{L\rightarrow \infty }\frac{\mathrm{ln}|\lambda _{2}|}{L}.
\label{the Lyapunov exponent of the edge states}
\end{equation}

In order to determine $\eta _{1}$, we perform a similarity transformation $%
ST_{j}S^{-1}=\xi ^{-1/2}\tilde{T}_{j}$ with $S=\mathrm{diag}(\xi ^{1/4},\xi
^{-1/4})$ and
\[
\xi =\frac{t_{3}-\gamma _{3}}{t_{2}-\gamma _{2}}.
\]%
After the similarity transformation, the transfer matrix $T_{j}$ of the
non-Hermitian AB model modulated by disorder is transformed into
\[
\tilde{T}_{j}=\left(
\begin{array}{cc}
\frac{V_{j}\sqrt{\xi }}{t_{3}-\gamma _{3}} & 1 \\
1 & 0%
\end{array}%
\right),
\]%
which is a transfer matrix of the Hermitian AA lattice. However, the
non-Hermiticity and nontrivial topological properties of the model are
transformed into the disorder term and an extra parameter $\xi ^{-1/2}$. The
transfer matrix of the whole system is given by $T=\Pi _{j=1}^{L}T_{j}=\xi
^{-L/2}\Pi _{j=1}^{L}\tilde{T}_{j}=\xi ^{-L/2}\tilde{T}$. Due to the
deficiency of the edge state in the Hermitian AA lattice, the Lyapunov
exponent of the model corresponding to the transfer matrix $\tilde{T}_{j}$
is zero obviously which lead to the Lyapunov exponent of the edge state $%
\eta _{1}=\mathrm{ln}\xi ^{-1/2}$ from the Eq. (\ref{the Lyapunov exponent
of the edge states}).

The second Lyapunov exponent concerns the Anderson localization of all
the states. According to the Thouless mechanism, the Lyapunov exponent of the
Anderson localized state in the neighborhood of $E_{B}$ can be computed as
\cite{Thouless_1972}
\[
\eta _{2}=\int d\varepsilon \rho \left( \varepsilon \right) \left[ \ln
\left\vert \varepsilon -E_{B}\right\vert -\mathrm{\ln }\left( \epsilon
_{+}\right) -\mathrm{\ln }\left( \epsilon _{-}\right) \right] ,
\]%
where $\rho (\varepsilon )$ is the density of state. The parameters $%
\epsilon _{+}$ and $\epsilon _{-}$ are introduced to reset the energy scale.
$E_{B}$ is set to be 0 for the sake of simplicity since $\eta _{2}$ is
energy-independent according to numerical analysis. $\eta _{2}$ is used to
characterize the bulk localization. The following derivations of $\eta _{2}$
of the model %in Eq. (\ref%
%{the block off-diagonal Hamiltonian})
are based on the Ref. \cite{PhysRevB.103.014201}. Under the OBC, we
therefore obtain the Lyapunov exponent $\eta _{2,\pm }=\mathrm{ln}\frac{%
Ve^{\alpha }}{\epsilon _{\pm }}$ where $\epsilon _{\pm }=\sqrt{\left(
t_{2}\pm \gamma _{2}\right) \left( t_{3}\pm \gamma _{3}\right) }$. Since the
log function needs $Ve^{\alpha }>\epsilon _{\pm }$, we can get two
localization transition points $V_{\text{L},\pm }=e^{-\alpha }\epsilon _{\pm
}$. $\eta _{2,+}$ is meaningless since the system has been in Anderson
localization phase when $V>V_{\text{L},-}=e^{-\alpha }\epsilon _{-}$. When $%
V>V_{\text{L},+}=e^{-\alpha }\epsilon _{+}$, the large quasiperiodic intensity
modifies the decay exponent of the Anderson localized state according the
formulae $\ \mathrm{exp}\left( -\eta \left\vert n-n_{0}\right\vert \right) $%
. So the Lyapunov exponent is given by $\eta _{2,-}=e^{-\alpha }\epsilon
_{-} $ which indicates that the onsite non-Hermiticity $\alpha $ and the
nonreciprocal parameters ($\gamma _{2}$, $\gamma _{3}$) are all negative
correlation with the Anderson transition.

When the localization of the nontrivial topological state is further
modulated by the Anderson localization of all the states, it gives the third
Lyapunov exponent. Combining the boundary localization of the nontrivial state
and bulk localization, the Lyapunov exponent is given by
\begin{equation}
\eta =\eta _{2,-}-\eta _{1}=\mathrm{ln}\left( Ve^{\alpha }-\tau _{-}\right) .
\label{The Lyapunov exponent in high V region}
\end{equation}%
As a result, the transition point of the nontrivial topological state is $V=e^{-\alpha }\tau _{-}$
which equals to the topological transition point $V_{\text{T},-}$.

Under the PBC, the non-Hermitian skin effect vanishes even in the presence
of the nonreciprocal hopping. In the absence of the asymmetrical
localization, the calculation of Lyapunov exponent can consult the Ref. \cite%
{PhysRevB.103.214202}. We can obtain two values of the fourth Lyapunov
exponent $\eta_\pm =\mathrm{ln}\frac{Ve^{\alpha}}{\tau _{\pm }}$ which are
the same as the winding numbers. Since the log function needs $%
Ve^{\alpha}>\tau _{\pm }$, we can get two localization transition points $%
V=e^{-\alpha}\tau _{\pm }$ which are just the topological transition points $%
V_{\text{T},\pm}$.%
% which also indicate that the onsite non-Hermiticity $\alpha$ is
%detrimental to localization transitions.

\section{Phase diagram}

\label{Phase diagram}

With the topological transition points $V_{\text{T},\pm }$ and the Anderson
transition points $V_{\text{L},\pm }$ in analysis forms, we are ready to concretely discuss
the two kinds of transitions and their relations of the model. It is no different
that whether the inverse of the golden ratio or
rational numbers $F_{n}/F_{n+1}$ is taken to be as $\beta$. When discussing the
topological Anderson insulator under the OBC, the nontrivial zero-edge
states are protected by the symmetry and the Anderson localized states
determined by the disorder intensity, provided $L$ is long enough. Under the PBC, the domain wall forms when bending the chain into a ring. The domain states are
protected by the symmetry and can be characterized by the winding number \cite{PhysRevB.100.035102, Han_2021}.
In particular, the Anderson localized states are from all the states of the ring and the
contribution of the domain states is negligible. In addition, an open boundary can be taken as a special domain wall. As we know, the periodic
boundary spectra can be well approximated by the open boundary spectra \cite{PhysRevLett.125.226402, Han_2021}. The argument also supports the fact that we can obtain the same conclusion under two boundary conditions. As the result, we take the
quasiperiodic parameter to be $\beta =(\sqrt{5}-1)/2$ directly in numerical analysis.
 The metric is
the chain limitation $L=F_{n+1}$ can be discarded.

\begin{figure}[tbp]
\begin{center}
\hspace*{-0.2cm} \includegraphics[scale=1]{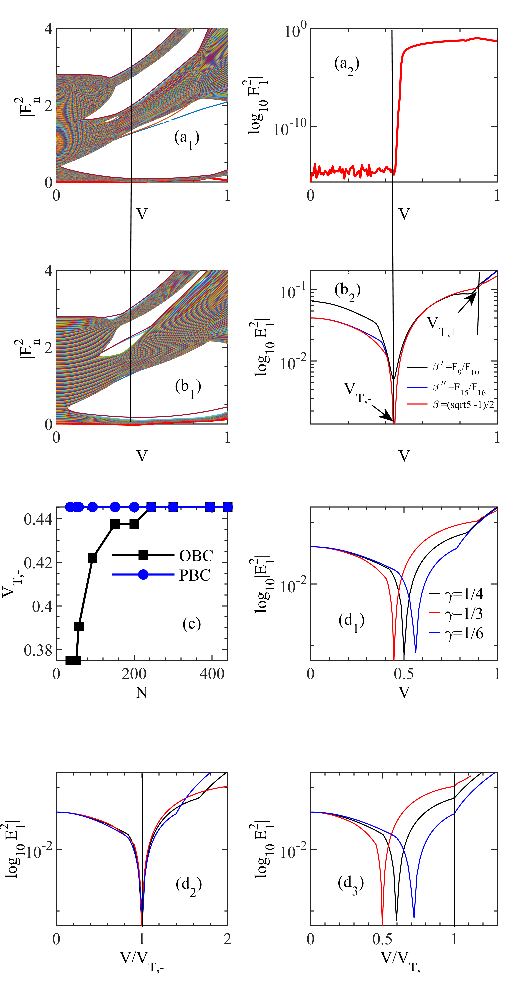}
\end{center}
\caption{Absolute values of spectra $|E_{n}^{2}|$ versus $V$ for the
non-Hermitian SSH model in Eq. (\protect\ref{the eigenvalue problem}) under
the OBC (a$_{1}$) and the PBC (b$_{1}$). The parameters are taken to be $%
t_{2}=1$, $t_{3}=0.8$, $\protect\alpha =0.4$ and $\protect\gamma =1/3$. The lattice
length is taken to be $L=1200$. (a$_{2}$)
The semilog of the lowest values of $|E_{n}^{2}|$ in (a$_{1}$)%
. (b$_{2}$) The semilog plots of the lowest values of $E_{n}$ under the
PBC with different $\beta$ and $L$. Black: $\beta =F_{9}/F_{10}, L =F_{10} =55$. Blue: $\beta =F_{15}/F_{16}, L =F_{16} =987$ and Red: $\beta  =(\sqrt{5}-1)/2$, $L =1200$. (c) The finite-size analysis of the transition point $V_{\text{T},-}$ under the
OBC [black square solid line] and PBC [blue point solid line] (d$_{1}$)
Semilog plots of the bulk energy gap of PBC spectra versus $V$ with
the different $\protect\gamma $. Red: $\protect\gamma =1/3$, Blue: $\protect%
\gamma =1/6$ and Black: $\protect\gamma =1/4$. The other parameters remain
unchanged. (d$_{2}$) Semilog plot of the bulk energy gap versus $%
V/V_{\text{T},-}$. (d$_{3}$) Semilog plot of the bulk energy gap versus $%
V/V_{\text{T},+}$. }
\label{energy_spectra}
\end{figure}

The OBC spectra shown in Fig. \ref{energy_spectra} (a$_{1}$) and PBC spectra
in Fig. (b$_{1}$) respectively. In the numerical analysis, $t_{2}$ is set to
be a unit parameter and other parameters are $t_{3}=0.8$, $\alpha =0.4$ and $%
\gamma _{2}=\gamma _{3}=\gamma =1/3$. %According to
%the Eq. (\ref{The parameter tau_pm}), $\tau _{\pm }=t_{2}\pm \gamma $.
Comparing the two energy spectra, the zero-energy states are found in the
OBC and PBC chains. To estimate the topological phase transitions, we plot
semilog of the lowest $|E_{n}^{2}|$ of the two spectra in Fig. \ref%
{energy_spectra} (a$_{2}$) and (b$_{2}$) under the two kinds of boundary
conditions. It finds that the open/close of the gap in Fig. \ref%
{energy_spectra} (b$_{2}$) is accompanied by the emerging of zero-energy in
Fig. \ref{energy_spectra} (a$_{2}$). Hence, the dip of the energy in Fig. %
\ref{energy_spectra} (b$_{2}$) can be taken as an indicator to characterize
the topological phase transition of the model.
The first transition point in Fig. \ref{energy_spectra} (a$_{1}$) is
consistent with the prediction by the formula $V_{\text{T},-}=\tau
_{-}e^{-\alpha}=\left( t_{2}-\gamma \right) e^{-\alpha}=0.4469$. However,
the second transition point $V_{\text{T},+}=\tau _{+}e^{-\alpha}=\left(
t_{2}+\gamma \right) e^{-\alpha}=0.8938$ don't accompany the gap closing of
the OBC spectra in Fig. \ref{energy_spectra} (a$_{1}$), and the small dip
appears in the PBC energy spectra in Fig. \ref{energy_spectra} (b$_{2}$). As previously pointed out, with increasing the disorder intensity $V$ across $V_{\text{T},-}$,
the nontrivial edge state is destroyed. Further increasing the disorder
intensity $V$ across $V_{\text{T},+}$, the system has been in the
topological trivial phase. As a result, the topological transition point is $%
V_{\text{T},-}$. We will see that the dip near $V_{\text{T},+}$ is related
to the Anderson transition. %Comparing the Hermitian
%case $\gamma = 0$ and $\alpha = 0$, a large $t_2$ is needed to obtain the
%same transition point. We therefore conclude that the non-Hermiticity $%
%\gamma $ is detrimental to the topological transition.

We calculate the lattice length $L$ dependence of two transition points in
Fig. \ref{energy_spectra} (c) under the two kinds of boundary conditions.
It finds that the topological transition approaches $V_{\text{T},- }=\tau
_{- }e^{-\alpha }$ under the OBC when $L$ is larger than $900$. Under the
PBC however, the chain length $L>500$ is enough to ensure the calculation
accuracy. So $L$ is taken to be $1200$ in all the numerical calculations.
To verify the effectivity of the above numerical method, we also present the semilog plots of the lowest values of $E_{n}$ under the
PBC with different chain length: $L=F_{10}=55$ [black line] and $L=F_{16}=987$ [blue line] in Fig. \ref{energy_spectra} (b$_{2}$). Corresponding, $\beta$ are taken to be $F_{9}/F_{10}$ and $F_{15}/F_{16}$ respectively. In the short lattice length $L=F_{10}=55$ case, the error of the second transition point $V_{\text{T},+ }$ is obvious. With increasing the lattice length $L=F_{16}=987$,  all the two transition points are consistent with the case $\beta =(\sqrt{5}-1)/2$ and $L=1200$.

To further verify the topological transition of the model, we plot the semilog of
the lowest $|E_{n}^{2}|$ of the PBC spectra in Fig. \ref{energy_spectra} (d$%
_{1}$) with different parameters. The amplitude of the quasiperidoic
potential $V$ is scaled by $V_{\text{T},-}$ (d$_{2}$) and $V_{\text{T},+}$ (d%
$_{3}$) respectively. The two dips collapse to the points which indicate the
effectiveness of $V_{\text{T},\pm }=\tau _{\pm }e^{-\alpha }$ .

%It should emphases that the energy spectra can be discussed in GBZ when the
%parameter $\beta $ is approximated as an irrational number $F_{n}/F_{n+1}$
%and the period of the external potential is $F_{n+1}$-lattice. If the
%conventional BZ is used however, the PBC spectra characterized by
%conventional wave vector is different from OBC spectra due to the
%non-Hermitian skin effect. Here a close chain is used to calculate the
%energy spectra without involving the conventional wave vector, the PBC
%spectra is the same as the OBC spectra except for the zero energies.

\begin{figure}[tbp]
\begin{center}
\hspace*{-0.3cm} \includegraphics[scale=0.8]{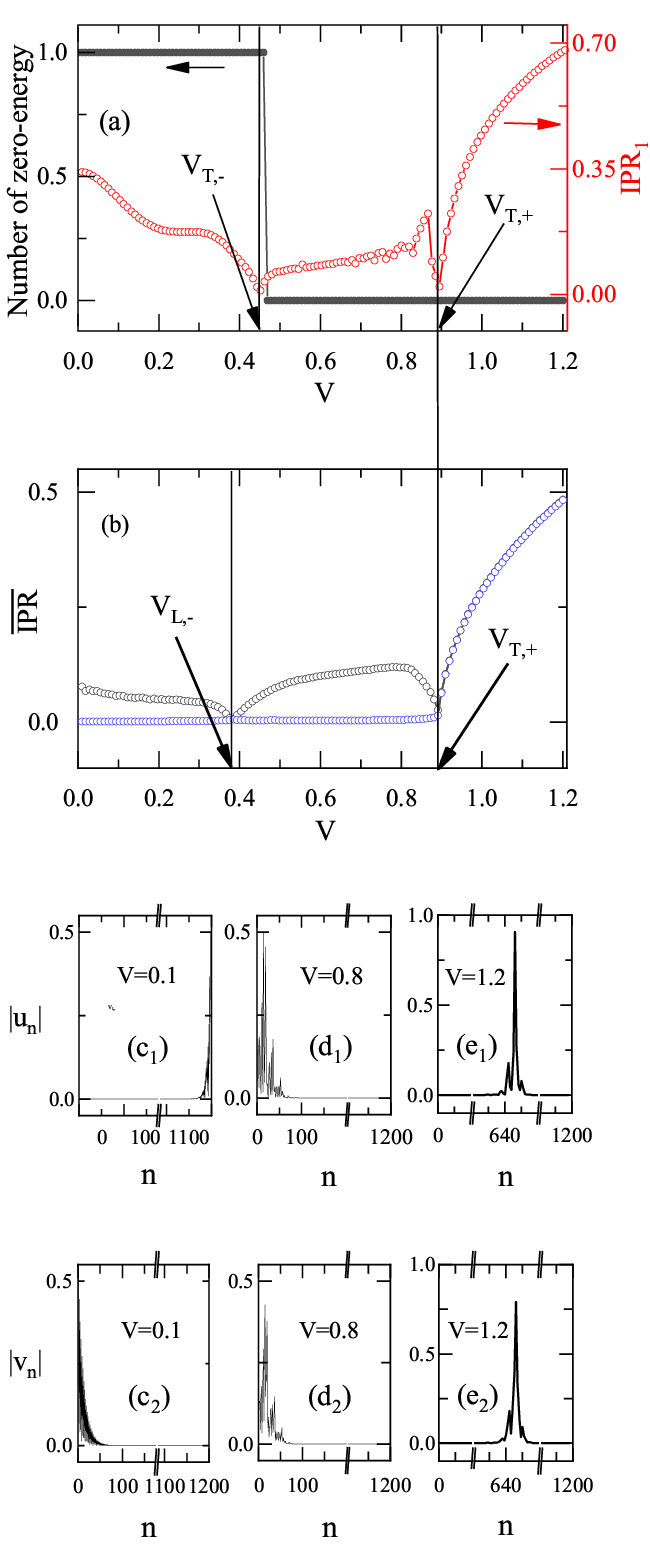}
\end{center}
\caption{Localization transitions. (a) $\text{IPR}_1$ and number of zero-energy state versus $V$. (b) $%
\overline{\text{IPR}}$ of the state $u$ versus $V$. The second row and the
third row are the wave-function $u_n$ and $v_n$ of the lowest $|E^2_n|$ for
the three different regions $V$ under the OBC. }
\label{phase_transition}
\end{figure}

The inverse of the participation ratio
\[
{\text{IPR}_{n}}=\sum\limits_{m=1}^{L}\left\vert u_{n}(x_m)\right\vert ^{4}
\]
can be used to characterize the localization of a normalized state $u_{n}(x_m)$.
For an extended state, $\text{IPR}_{n}$ is of the order $1/L$, whereas it
approaches to $1$ for a localized state. When $v_{n}$ is used to calculate $%
\text{IPR}_{n}$, the conclusions remain unchanged.
the localization of the whole system can be characterized by the mean inverse of the participation
ratio \[\overline{\text{IPR}}=\sum_{n}{\text{IPR}_{n}}/L.\]

We calculate $%
\text{IPR}_{1}$ corresponding to the lowest $|E_{n}^{2}|$ shown in Fig. \ref%
{phase_transition} (a) to understand the localization of the nontrivial topological phase. As a demonstration of the topological phase, we also
plot the zero-energy number versus $V$ in Fig. \ref{phase_transition} (a). In
the low $V$ region near $V_{\text{T},-}=e^{-\alpha }(t_{2}-\gamma )=0.4469$,
a dip $V\approx 0.45$ dives in Fig. \ref{phase_transition} (a). The
appearance of the dip (delocalization) accompanying the topological
transition is due to the localization competition between the nontrivial edge
states and the disorders. The competition leads to the most extended case of
the particle distributions (the smallest $\text{IPR}_{1}$) in the transition
point. As expected, the zero-energy edge states [Fig. \ref{phase_transition}
(c$_{1}$) and (c$_{2}$)] indicate the system in the topological nontrivial
phase when $V<V_{\text{T},-}$. When $V>V_{\text{T},-}$, the edge state
(nontrivial topological phase) is breakdown. Further increasing $V>V_{\text{T%
},+}=e^{-\alpha }(t_{2}+\gamma )=0.8938$, the edge state cannot exist.
However, the other dip appears at $V=0.89$ in Fig. \ref{phase_transition}
(a) and (b). As the result, the topological transition point is $V_{\text{T}%
,-}=e^{-\alpha }(t_{2}-\gamma )$. The consistence of the above numerical result with the formulae $V_{\text{T%
},-}$ verifies that the
topological transition accompanies the breakdown of the zero-energy
edge state, and the large non-Hermitian parameters $\alpha $ and $\gamma $
are all detrimental to the topological transition.
%The main prediction of the theory is $V_{A}/V_{T}=\alpha $%
%. From the Fig. \ref{energy_spectra} (b$_{2}$) (b$_{2}$) and Fig. \ref%
%{phase_transition} (b), the calculated topological transition is $V_{T}=0.45$%
%. From the first dip of $\overline{\text{IPRs}}$ in Fig. \ref%
%{phase_transition} (a), the calculated Anderson transition is $V_{A}=0.37$.
%The numerical $\alpha \approx 0.82$ is near the theoretical prediction 074.

We calculate the averaged inverse participation ratios ($\overline{\text{IPR}%
}$) over the right eigenstates $u_{n}$ in Eq. (\ref{the eigenvalue problem})
under two kinds of boundary condition in Fig. \ref{phase_transition} (b) to
understand the Anderson localization. In the low $V$ region near $V_{\text{L}%
,-}=e^{-\alpha }\epsilon _{-}=e^{-\alpha }\sqrt{(t_{2}-\gamma )(t_{3}-\gamma
)}=0.3739$, a deep $V\approx 0.38$ dives in Fig. \ref{phase_transition} (b)
under the OBC which is different from that in Fig. \ref{phase_transition}
(a). The appearance of the dip indicates the Anderson transition of the whole system
doesn't accompany the topological transition which is contrast to the case of $\text{IPR}_{1}$.

The inconsistence of the topological transition and Anderson transition
can be understood as follow.
When calculating $\overline{\text{IPR}}$,
the contribution of edge
zero-energy states is negligible. The delocalization transition is due to
the localization competition between boundary localization of the bulk states from the
non-Hermitian skin effect and the Anderson localization from the disorders. It is that non-Hermitian skin effect leads to all the states to localize at the boundary.
In the absence of the non-Hermitian skin effect, all the bulk states are Bloch waves and the nontrivial topological states localize at the boundary. The boundary localization is determined by the nontrivial topological states only, which the Anderson transition accompanies the topological transition.
$V_{\text{T}%
,-}>V_{\text{L}%
,-}$ also indicates that the topological edge states are protected by
the symmetry, even in the non-Hermtian case, and the topological phases are
expected to be immune to the perturbations of disorders.

Further increasing $V$ $(>e^{-\alpha }\epsilon _{+})$, the system has been
in the Anderson localization phase. We get the other transition point $V_{%
\text{T},+}=e^{-\alpha }\tau _{+}=0.89$. Across the dip, $\overline{\text{IPR%
}}$ increases remarkably and the system in the Anderson localization phase
shown in Fig. \ref{phase_transition} (e$_{1}$) and (e$_{2}$). In the
intermediate region, $V_{\text{L},-}<V<V_{\text{T},+}$, the spatial
distribution of wave functions is shown in Fig. \ref{phase_transition} (d$%
_{1}$) and (d$_{2}$).
%The localization is due to the Anderson localization and non-Hermitian skin effect.
This transition also occurs under the PBC where $\overline{\text{IPR}}$
increases rapidly across the transition point [blue-circle in Fig. \ref%
{phase_transition} (b)]. In such case, the non-Hermitian skin effect and
zero-energy edge states disappear. Therefore, there isn't the
boundary-localization nature. The Anderson localization is completely due to
the disorders which lead to the destructive interference of scattered waves.
The consistence of the second transition points in Fig. \ref%
{phase_transition} (a) and (b) indicates the delocalizations have the same
origin. Although it can be characterized by the winding number, the Anderson
localization is unrelated to the topological transition.
%As the result, the appearance of Anderson localization is due to
%the strong disorder only. So the transition has nothing to do with the
%topological transition and the disorders dominate the localization behavior.
According to the formula $V_{\text{T},+}=\left( t_{2}+\gamma \right)
e^{-\alpha }$, we conclude that the non-Hermitian skin effect enhances the
Anderson localization.

\section{Summary}

\label{summary}

In summary, we have studied the localization and topological phase
transitions of non-Hermitian SSH models, where the non-Hermiticities are
introduced by the complex quasiperiodic hopping and the nonreciprocal
hopping. In the presence of the nonreciprocal hopping, the induced
non-Hermitian skin effect leads to asymmetric localization of all the states.
Under the OBC, increasing the intensity of the complex quasiperiodic hopping
destroys the nontrivial zero-energy edge state and asymmetric localized
states, and drives all the states into Anderson localized states. The
competition between boundary localization and Anderson localization leads to
the delocalization transitions. Due to the non-Hermitian skin effect, the
localization transitions are not necessarily accompanied by the topological
transitions. By analysing the winding number of energy and the Lyapunov
exponent, we find the large nonreciprocal hopping is detrimental to the
topological phase transitions, and enhances the Anderson localization.
However, the large on-site non-Hermiticity is always detrimental to the
Anderson localization and topological phase transitions. Since the models
have many topological equivalent models, the results we studied are useful
for further studying the topological Anderson insulator experimentally.

\begin{acknowledgments}
This work was supported by Hebei Provincial Natural Science Foundation of
China (Grant No. A2012203174, No. A2015203387), Science and Technology Project of Hebei Education
Department, China (Grant No. ZD2020200) and National Natural Science
Foundation of China (Grant No. 10974169, No. 11304270).
\end{acknowledgments}

%\bibliography{Refs}

\end{document}